\documentclass[12pt]{article}

\def \g{\gamma}
\def \d{\delta}
\def \ep{\epsilon}

\def \n{\nu}

\def \s{\sigma}

\def \c{\chi}
\def \ps{\psi}

\def \L{\Lambda}

\def \Ps{\Psi}

\def \la#1{\label{#1}}

\def \le{\left}
\def \ri{\right}

\def \da{\dagger}

\def \lb{\lbrack}
\def \rb{\rbrack}

\def \rar{\rightarrow}
\def \uar{\uparrow}
\def \dar{\downarrow}
\def \lrar{\leftrightarrow}
\def \ld{\ldots}
\def \cd{\cdots}
\def \nn{\nonumber}

\newcommand \beq{\begin{eqnarray}}
\newcommand \eeq{\end{eqnarray}}
\newcommand \beqs{\begin{eqnarray}}
\newcommand \eeqs{\begin{eqnarray}}
\newcommand \ba{\begin{array}}
\newcommand \ea{\end{array}}

\def \ma{{\rm matrix}}

\begin{document}
\begin{center}
   {\LARGE\bf  Integrability of Supersymmetric Quantum Matrix Models in the Large-$N$ Limit} \\
   \vspace{2cm}
   {\large\bf C.-W. H. Lee and  S. G. Rajeev} \\
   {\it Department of Physics and Astronomy, University of Rochester, 
    Rochester, New York 14627} \\
   \vspace{.5cm}
   {June 1st, 1998} \\
   \vspace{2cm}
   {\large\bf Abstract}
\end{center}

Many physical systems like supersymmetric Yang-Mills theories are formulated as quantum matrix models.  We discuss 
how to apply the Bethe ansatz to exactly solve some supersymmetric quantum matrix models in the large-$N$ limit.  
Toy models are constructed out of the one-dimensional Hubbard and t-J models as illustrations.  

\begin{flushleft}
\end{flushleft}
\pagebreak

More and more physical systems are formulated as quantum matrix models; notable examples are quantum chromodynamics
\cite{thooft, witten, thorn79, rajeev, anda}, supersymmetric Yang-Mills theory \cite{kutasov, susyym}, D-branes 
\cite{dbrane, polchinski}, M-theory \cite{mtheory, taylor} and string theory \cite{thorn96, verlinde}.  The 
dimensions of these matrices can be interpreted as the numbers of colors or the numbers of D-branes.  Taking the 
limit as the dimensions going to infinity is also of interest; for instance, it is conjectured that the large-$N$ 
limit captures the essence of quantum chromodynamics like quark confinement \cite{thooft}, and that M-theory in the 
infinite momentum frame is exactly described by the large-$N$ limit of D0-brane quantum mechanics \cite{mtheory}.  
Certainly if ways are developed to solve quantum matrix models in the large-$N$ limit, we will understand better 
the physics of these interesting systems.

In this Letter, we will concentrate ourselves on matrix models whose Hamiltonians preserve the number of partons.  
These models can be used to describe asymptotic spectra of supersymmetric Yang-Mills theories \cite{kutasov}.  It 
is found that the ubiquitous Bethe ansatz \cite{bethe} is helpful in solving some of these models, and the 
Yang-Baxter equation can help us determine if the Bethe ansatz is applicable.  In previous papers \cite{prl, 
opstal}, we provided some examples of bosonic quantum matrix models which can be solved by the Bethe ansatz.  Here, 
we will give first instances of {\em quantum matrix models which contain both fermions and bosons and which are 
still exactly integrable}.  Previously, work was done to solve some {\em integrals} over finite chains of matrices
with both commuting and anti-commuting entries which describe {\em classical} integrable systems with 
supersymmetry (see Ref.\cite{semenoffszabo} and the citations therein).  From this perspective, the implication of 
our discussion here is that certain {\em path integrals} over matrices corresponding to {\em quantum} integrable 
systems with supersymmetry can be exactly evaluated, too. 

Consider a physical system of bosons with $\L_B$ degrees of freedom (other than colors or labels for D-branes) and 
fermions with $\L_F$ degrees of freedom (again other than colors).  (In the context of D-branes, $\L_B$ is the
number of transverse dimensions of the D-branes; in the context of gauge theory, $\L_B$ gives the number of 
different values of the longitudinal momentum a gluon can take.)   Let the vacuum be trivial.  Let 
$a^{\mu_1}_{\mu_2}(k)$ be an annihilation operator for a boson if $1 \leq k \leq \L_B$, or a fermion if 
$\L_B + 1 \leq k \leq \L_B + \L_F$.  Here $\mu_1$ and $\mu_2$ are row and column indices running from 1 to $N$ 
(physically they can be color indices or labels for D-branes).  The corresponding creation operator is 
$a^{\da\mu_1}_{\mu_2}(k)$.  They satisfy typical (anti)-commutation relations, the most important ones being
\[ \le\lb a^{\mu_1}_{\mu_2}(k_1), a^{\da\mu_3}_{\mu_4}(k_2) \ri\rb = 
   \d_{k_1 k_2} \d^{\mu_3}_{\mu_2} \d^{\mu_1}_{\mu_4} \]
for $1 \leq k_1, k_2 \leq \L_B$; and
\[ \le\lb a^{\mu_1}_{\mu_2}(k_1), a^{\da\mu_3}_{\mu_4}(k_2) \ri\rb_+ = 
   \d_{k_1 k_2} \d^{\mu_3}_{\mu_2} \d^{\mu_1}_{\mu_4} \]
for $\L_B + 1 \leq k_1, k_2 \leq \L_B + \L_F$.  

A typical color-invariant physical state is a linear combination of states of the form
\beq 
   \Ps^K \equiv N^{-c/2} a^{\da\n_2}_{\n_1}(k_1) a^{\da\n_3}_{\n_2}(k_2) \cd a^{\da\n_1}_{\n_c}(k_c) |0 \rangle.
\la{1}
\eeq
Here we sum over all possible values of the row and column indices.  The superscript $K$ is the integer sequence 
$k_1$, $k_2$, \ld, $k_c$.  The factor of $N$ in this equation serves as a normalization factor in the large-$N$ 
limit.  If we split $K$ into two sequences $K_1$ and $K_2$ such that $K_1 K_2 = K$, it follows that
\beq 
   \Ps^K = (-1)^{\ep(K_1) \ep(K_2)} \Ps^{K_2 K_1} 
\la{1.1}
\eeq
where $\ep(K_1) = 0$ if the number of integers in $K_1$ larger than $\L_B$ is even, and $\ep(K_1) = 1$ if it is 
odd. 

A color-invariant observable is a linear combination of operators of the form
\begin{eqnarray*}
   \g^I_J & \equiv & N^{-(a -1)} a^{\da\mu_2}_{\mu_1}(i_1) a^{\da\mu_3}_{\mu_2}(i_2) \cd 
   a^{\da\n_a}_{\mu_a}(i_a) \cdot \\
   & & a^{\n_{a-1}}_{\n_a}(j_a) a^{\n_{a-2}}_{\n_{a-1}}(j_{a-1}) \cd a^{\mu_1}_{\n_1}(j_1).
\end{eqnarray*}
Again we sum over all possible values of the row and column indices. $I$ is the integer sequence $i_1$, $i_2$, \ld, 
$i_a$, and $J$ is the sequence $j_1$, $j_2$, \ld, $j_a$.  The factor of $N$ is chosen to ensure that once we act 
this operator on a state of the form given by Eq.(\ref{1}), we will get another normalizable state.

It is known that as $N \rar \infty$, if there were no fermions, the action of $\g^I_J$ on $\Ps^K$ would be such 
that a segment of any cyclic permutation of $K$ which is identical to $J$ would be replaced with $I$ \cite{thorn79, 
lee}.  This is the key reason why there is a connection between large-$N$ matrix models and spin chains.  With the 
inclusion of fermions, the situation is similar except that we have to be careful with the signs of the terms:
\beq
   \g^I_J \Ps^K & = & \d^K_J \Ps^I + \sum_{K_1 K_2 = K} (-1)^{\ep(K_1) \ep(K_2)} \d^{K_2 K_1}_J \Ps^I
   + \sum_{K_1 K_2 = K} \d^{K_1}_J \Ps^{I K_2} \nn \\
   & & + \sum_{K_1 K_2 K_3 = K} (-1)^{\ep(K_1) \lb \ep(K_2) + \ep(K_3) \rb} \d^{K_2}_J \Ps^{I K_3 K_1} \nn \\
   & & + \sum_{K_1 K_2 = K} (-1)^{\ep(K_1) \ep(K_2)} \d^{K_2}_J \Ps^{I K_1} \nn \\
   & & + \sum_{J_1 J_2 = J} \sum_{K_1 K_2 K_3 = K} (-1)^{\ep(K_3) \lb \ep(K_1) + \ep(K_2) \rb}
   \d^{K_3}_{J_1} \d^{K_1}_{J_2} \Ps^{I K_2}.
\la{3}
\eeq
In Eq.(\ref{3}), a summation like $\sum_{K_1 K_2 = K}$ means that we sum over all possible pairs of $K_1$ and 
$K_2$ such that the concatenated sequence is $K$.  A typical delta function like $\d^K_J$ yields 1 if the sequence
$K$ is the same $J$, and 0 otherwise.  

A crucial observation at this point is that Eq.(\ref{3}), an equation in the context of quantum matrix models in 
the large-$N$ limit, can be paraphrased in the context of a quantum spin chain containing both bosons and fermions 
as follows.  Consider a spin chain with $c$ sites satisfying the periodic boundary condition.  Each site can be 
occupied either by a boson or a fermion.  There are $\L_B$ possible bosonic states, and $\L_F$ possible fermionic
ones.  Let $A_p(1)$, $A_p(2)$, \ld, $A_p(\L_B)$ be the annihilation operators of these bosonic states, and
$A_p(\L_B + 1)$, $A_p(\L_B + 2)$, \ld, $A_p(\L_B + \L_F)$ be the annihilation operators of the fermionic ones.
Attaching daggers as superscripts to these operators turns them to the corresponding creation operators.  Then a 
typical state of a spin chain can be written as
\[ \ps^K \equiv A^{\da}_1(k_1) A^{\da}_2(k_2) \cd A^{\da}_c(k_c) |0 \rangle. \]
Define the {\em Hubbard operator} \cite{bablog} as follows:
\[ X^{ij}_p \equiv A^{\da}_p(i) A_p(j). \] 
It is readily seen that the action of a product of Hubbard operators on $\ps^K$ is
\beq
   \lefteqn{X^{i_1 j_1}_p X^{i_2 j_2}_{p+1} \cd X^{i_a j_a}_{p+a-1} \ps^K = 
   (-1)^{\lb \ep(k_1) + \ep(k_2) + \cd + \ep(k_{p-1}) \rb \lb \ep(I) + \ep(J) \rb } \cdot} \nn \\
   & & (-1)^{\lb \ep(i_a) + \ep(j_a) \rb \lb \ep(k_p) + \ep(k_{p+1}) + \cd + \ep(k_{p+a-2}) \rb} 
   \d^{k_{p+a-1}}_{j_a} \cdot \nn \\
   & & (-1)^{\lb \ep(i_{a-1}) + \ep(j_{a-1}) \rb \lb \ep(k_p) + \ep(k_{p+1}) + \cd + \ep(k_{p+a-3}) \rb}
   \d^{k_{p+a-2}}_{j_{a-1}} \cd \nn \\
   & & (-1)^{\lb \ep(i_2) + \ep(j_2) \rb \ep(k_p)} \d^{k_{p+1}}_{j_2} \d^{k_p}_{j_1}
   \ps^{k_1 k_2 \ld k_{p-1} i_1 i_2 \ld i_a k_{p+a} k_{p+a+1} \ld k_c}
\la{4}
\eeq
for $p+a-1 \leq c$.

We are now ready to identify supersymmetric matrix models with supersymmetric quantum spin chains.  Consider the
following weighted sum of all cyclic permutations of the states of a quantum spin chain in the following way:
\[ \Ps^K = \sum_{K_1 \dot{K}_2 = K} (-1)^{\ep(K_1) \ep(\dot{K}_2)} \ps^{\dot{K}_2 K_1} \]
where $\dot{K}_2$ may or may not be empty but $K_1$ has to be non-empty.  This $\Ps^K$ obeys Eq.(\ref{1.1}), too.  
Then after some algebra, it can be shown that
\beq
   \g^I_J & = & (-1)^{\lb \ep(i_a) + \ep(j_a) \rb \lb \ep(j_1) + \ep(j_2) + \cd + \ep(j_{a-1}) \rb } \cdot \nn \\
   & & (-1)^{\lb \ep(i_{a-1}) + \ep(j_{a-1}) \rb \lb \ep(j_1) + \ep(j_2) + \cd + \ep(j_{a-2}) \rb } \cd
   (-1)^{\lb \ep(i_2) + \ep(j_2) \rb \ep(j_1)} \cdot \nn \\
   & & \sum_{p=1}^c X^{i_1 j_1}_p X^{i_2 j_2}_{p+1} \cd X^{i_a j_a}_{p+a-1}
\la{6}
\eeq
because both the left and right hand sides of it satisfies Eq.(\ref{3}).  As the Hamiltonian of a typical matrix
model is of the form $\sum_{I, J} h^J_I \g^I_J$, where only a finite number of $h^J_I$'s are non-zero, Eq.(\ref{6})
provide us a way to transcribe the matrix model into a quantum spin chain with fermions and/or bosons, if for
each non-zero $h^J_I$ the numbers of integers in $I$ and $J$ are the same.

Let us give some typical examples to illustrate the idea above.  Consider the following supersymmetric matrix
model with $\L_B = \L_F = 2$:
\begin{eqnarray*}
   H^{\ma}_{\rm Hubbard} & = & - \le( \g^{34}_{12} + \g^{34}_{21} - \g^{43}_{12} - \g^{43}_{21} 
   + \g^{12}_{34} + \g^{21}_{34} - \g^{12}_{43} - \g^{21}_{43} \ri) \\
   & & - \le( \g^{13}_{31} + \g^{31}_{13} + \g^{14}_{41} + \g^{41}_{14} \ri) +
   \le( \g^{23}_{32} + \g^{32}_{23} + \g^{24}_{42} + \g^{42}_{24} \ri)\\
   & & + U \g^2_2.
\end{eqnarray*} 
Eq.(\ref{6}) tells us that the corresponding spin chain model is
\beq
   H_{\rm Hubbard} & = & - \sum_{p=1}^n \sum_{\s = 3, 4} \le( X^{\s 1}_p X^{1 \s}_{p+1} - 
   X^{1 \s}_p X^{\s 1}_{p+1} \ri) \nn \\
   & & - \sum_{p=1}^n \le( X^{31}_p X^{42}_{p+1} + X^{24}_p X^{13}_{p+1}
   + X^{14}_p X^{23}_{p+1} + X^{32}_p X^{41}_{p+1} \ri. \nn \\
   & & \le. - X^{42}_p X^{31}_{p+1} - X^{13}_p X^{24}_{p+1} - X^{23}_p X^{14}_{p+1} - X^{41}_p X^{32}_{p+1} \ri) 
   \nn \\
   & & - \sum_{p=1}^n \sum_{\s = 3, 4} \le( X^{2 \s}_p X^{\s 2}_{p+1} - X^{\s 2}_p X^{2 \s}_{p+1} \ri) \nn \\
   & & + U \sum_{p=1}^n X^{22}_p.
\la{8}
\eeq
If we identify the states 1, 2, 3, 4 to be the vacuum state, the state with one spin-up and one spin-down 
electrons, the state with a spin-up electron only, and the state with a spin-down electron only respectively, we 
will see that the Hamiltonian in Eq.(\ref{8}) is nothing but that of the Hubbard model \cite{hubbard}.  More 
specifically, let $c_{p, \uar}$ and $c_{p, \dar}$ be the annihilation operators of spin-up and spin-down electrons 
at the $p$-th site respectively.  Then we can make the following identifications:
\[ c_{p, \uar} \lrar A_p(3); \; c_{p, \dar} \lrar A_p(4) \mbox{; and} \; c_{p, \dar} c_{p, \uar} \lrar A_p(2). \]
The Hamiltonian can now be rewritten as
\[ H_{\rm Hubbard} = - \sum_{p=1}^n \sum_{\s = \uar, \dar} \le( c^{\da}_{p\s} c_{p+1, \s} + 
   c^{\da}_{p+1, \s} c_{p\s} \ri) + U \sum_{p=1}^n n_{p\uar} n_{p\dar}. \]
where $n_{p\uar} = c^{\da}_{p\uar} c_{p\uar}$ and $n_{p\dar} = c^{\da}_{p\dar} c_{p\dar}$ are the number operators 
for spin-up and spin-down states at the $i$-th site.  This model was shown by Lieb and Wu to be exactly integrable
using the Bethe ansatz \cite{liwu}, and is believed to describe many condensed matter phenomena like high-$T_c$
superconductivity, fractional quantum Hall effect, superfluidity, ... etc. \cite{proceedings}

It is not necessary for $\L_B = \L_F$ for a matrix model to be integrable.  Consider another celebrated fermionic
spin chain model, the t-J model \cite{tj, bablog}.  Its Hamiltonian is given by
\begin{eqnarray*} 
   H_{\rm tJ} & = & -t \sum_{p=1}^n \sum_{\s = \uar, \dar} P \le( c^{\da}_{p\s} c_{p+1, \s} +
   c^{\da}_{p+1, \s} c_{p\s} \ri) P \nn \\
   & & + \frac{J}{2} \sum_{p=1}^n \le[ {\bf S}_p \cdot {\bf S}_{p+1} - 
   \frac{1}{4} \le( n_{p\uar} + n_{p\dar} \ri) \le( n_{p+1, \uar} n_{p+1, \dar} \ri) \ri].
\end{eqnarray*}
In this equation,
\[ P = \prod_{p=1}^n (1 - n_{p\uar} n_{p\dar}) \]
is the projection operator to the collective state in which no site has 2 electrons.  ${\bf S}_p$ is the usual spin
operator:
\begin{eqnarray*}
   S^x_p & = & \frac{1}{2} \le( S_p + S^{\da}_p \ri); \\
   S^y_p & = & \frac{1}{2 {\rm i}} \le( S_p - S^{\da}_p \ri); \\
   S^z_p & = & \frac{1}{2} \le( n_{p\uar} - n_{p\dar} \ri); \\
   S^{\da}_p & = & c^{\da}_{p\dar} c_{p\uar}; \; \mbox{and} \\
   S_p & = & c^{\da}_{p\uar} c_{p\dar}.  
\end{eqnarray*}
If $J \ll t$, then this model is equivalent to the Hubbard model in the large-$U$ limit.  However, this model is in 
general different from the Hubbard model.  When the values of $J$ and $t$ are such that $J = 2t$, we say that the 
model is at the supersymmetric point, and it is integrable.  The corresponding integrable matrix model in the 
large-$N$ limit is
\[ H^{\ma}_{\rm tJ} = -t \le( \g^{13}_{31} + \g^{31}_{13} + \g^{14}_{41} + 
   \g^{41}_{14} \ri) - \frac{J}{4} \le( \g^{34}_{43} + \g^{43}_{34} - \g^{34}_{34} - \g^{43}_{43} \ri). \]
Note that there are 2 fermionic states (States 3 and 4) but only 1 bosonic state (State 1) in this integrable model.

To date many exactly integrable generalizations of the Hubbard model have been found \cite{other}.  We can use the 
same transcription procedure to write down the corresponding integrable supersymmetric matrix models in the 
large-$N$ limit.

Besides considering matrix models in which there are only matrix degrees of freedom, we can also consider models,
the matrix-vector models, in which there are both matrix and {\em vector} degrees of freedom.  Such models 
describe, say, mesons in quantum chromodynamics or both quarks and squarks in supersymmetric Yang-Mills theory.  
For the sake of simplicity, we will confine ourselves to models in which there are only one column vector and one 
row vector degree of freedom.  Let $\c^{\mu}$ and $\bar{\c}_{\mu}$, where again $\mu$ runs from 1 to $N$, be 
annihilation operators such that typical commutation relations hold.  In particular, they obey the following two 
non-trivial commutators:
\beq
   \le\lb \c^{\mu_1}, \c^{\da}_{\mu_2} \ri\rb = \d^{\mu_1}_{\mu_2} \mbox{; and}
   \le\lb \bar{\c}_{\mu_1}, \bar{\c}^{\da\mu_2} \ri\rb = \d^{\mu_2}_{\mu_1}.
\la{9}
\eeq
The following formalism will still hold if we replace the commutators among $\c^{\mu}$'s and $\bar{\c}_{\mu}$'s 
with anti-commutators.  A typical color-invariant state is a linear combination of the form
\[ s^K \equiv N^{-(c+1)/2} \bar{\c}^{\da\n_1} a^{\da\n_2}_{\n_1}(k_1) a^{\da\n_3}_{\n_2}(k_2) \cd
              a^{\da\n_{c+1}}_{n_c}(k_c) \c^{\da}_{\n_{c+1}} | 0 \rangle. \]
An observable is again a linear combination of $\g^I_J$ introduced above.  Nevertheless, we find it convenient to
introduce the following two kinds of operators:
\begin{eqnarray*}
   l^I_J & \equiv & N^{-a} \bar{\c}^{\da\mu_1} a^{\da\mu_2}_{\mu_1}(i_1) a^{\da\mu_3}_{\mu_2}(i_2) \cd
   a^{\da\mu_{a+1}}_{\mu_a}(i_a) \cdot \\
   & & a^{\n_a}_{\mu_{a+1}}(j_a) a^{\n_{a-1}}_{\n_a}(j_{a-1}) \cd a^{\n_1}_{\n_2}(j_1) \bar{\c}_{\n_1}
\end{eqnarray*}
and
\begin{eqnarray*}
   r^I_J & \equiv & N^{-a} \c^{\da}_{\mu_{a+1}} a^{\da\mu_{a+1}}_{\mu_a}(i_a) 
   a^{\da\mu_a}_{\mu_{a-1}}(i_{a-1}) \cd a^{\da\mu_2}_{\mu_1}(i_1) \cdot \\
   & & a^{\mu_1}_{\n_1}(j_1) a^{\n_1}_{\n_2}(j_2) \cd a^{\n_{a-1}}_{\n_b}(j_a) \c^{\n_a}.  
\end{eqnarray*}
Actually an $l^I_J$ is a finite linear combination of $\g^I_J$'s, and so is an $r^I_J$, though it is no need for
us to bother about the details here.  (See Ref.\cite{sustal} for a comprehensive discussion of this.)  

Matrix-vector models correspond to spin chain models satisfying {\em open} boundary conditions.  The relation 
between a $\g$ and Hubbard operators is almost the same as Eq.(\ref{6}), except that the summation index $p$ runs
from 1 to $c - a + 1$ this time.  It can be easily seen that the corresponding Hubbard operators for 
an $l^I_J$ and an $r^I_J$ are
\begin{eqnarray*}
   l^I_J & = & (-1)^{\lb \ep(i_a) + \ep(j_a) \rb \lb \ep(j_1) + \ep(j_2) + \cd + \ep(j_{a-1}) \rb } \cdot \nn \\
   & & (-1)^{\lb \ep(i_{a-1}) + \ep(j_{a-1}) \rb \lb \ep(j_1) + \ep(j_2) + \cd + \ep(j_{a-2}) \rb } \cd
   (-1)^{\lb \ep(i_2) + \ep(j_2) \rb \ep(j_1)} \cdot \nn \\
   & & X^{i_1 j_1}_1 X^{i_2 j_2}_2 \cd X^{i_a j_a}_a
\end{eqnarray*}
and
\begin{eqnarray*}
   r^I_J & = & (-1)^{\lb \ep(i_a) + \ep(j_a) \rb \lb \ep(j_1) + \ep(j_2) + \cd + \ep(j_{a-1}) \rb } \cdot \nn \\
   & & (-1)^{\lb \ep(i_{a-1}) + \ep(j_{a-1}) \rb \lb \ep(j_1) + \ep(j_2) + \cd + \ep(j_{a-2}) \rb } \cd
   (-1)^{\lb \ep(i_2) + \ep(j_2) \rb \ep(j_1)} \cdot \nn \\
   & & X^{i_1 j_1}_{n-a+1} X^{i_2 j_2}_{n-a+2} \cd X^{i_a j_a}_n
\end{eqnarray*}
respectively.

Let us see how the transcription is put into practice.  Consider a matrix-vector model with the following 
Hamiltonian:
\begin{eqnarray*}
   H^{\ma}_{\rm Hubbard(o)} & = & - \le( \g^{34}_{12} + \g^{34}_{21} - \g^{43}_{12} - \g^{43}_{21} 
   + \g^{12}_{34} + \g^{21}_{34} - \g^{12}_{43} - \g^{21}_{43} \ri) \\
   & & - \le( \g^{13}_{31} + \g^{31}_{13} + \g^{14}_{41} + \g^{41}_{14} \ri) +
   \le( \g^{23}_{32} + \g^{32}_{23} + \g^{24}_{42} + \g^{42}_{24} \ri) \\
   & & - U \g^2_2 + \mu \le( \g^3_3 + \g^4_4 + 2 \g^2_2 \ri) \\
   & & - p_{\uar} \le( l^3_3 + r^3_3 + l^2_2 + r^2_2 \ri) - p_{\dar} \le( l^4_4 + r^4_4 + l^2_2 + r^2_2 \ri).
\end{eqnarray*}
It turns out that the corresponding spin chain model is nothing but a Hubbard model with an open boundary condition
\cite{schultz, assu, zhou}:
\begin{eqnarray*}
   H_{\rm Hubbard(o)} & = & - \sum_{i=1}^{n-1} \sum_{\s = \uar, \dar} 
   \le( c^{\da}_{i\s} c_{i+1, \s} + c^{\da}_{i+1, \s} c_{i\s} \ri) + U \sum_{i=1}^n n_{i\uar} n_{i\dar} \nn \\
   & & + \mu \sum_{i=1}^n \le( n_{i\uar} + n_{i\dar} \ri) - p_{\uar} \le( n_{1\uar} + n_{n\uar} \ri)
   - p_{\dar} \le( n_{1\dar} + n_{n\dar} \ri).
\end{eqnarray*}
(We remark that this particular Hubbard model is taken from Ref.\cite{assu}.)

We have seen in the above discussion how to determine the integrability of a supersymmetric quantum matrix model in
the large-$N$ limit --- rewrite the Hamiltonian in terms of Hubbard operators, and then test if the Bethe ansatz is
applicable by checking if the scattering matrix of the equivalent spin chain model satisfies the Yang-Baxter 
equation.  As for the case of bosonic spin chains, we remark that the operators $\g$, $l$ and $r$ defined above 
form a Lie superalgebra \cite{sustal}.  The symmetry of the simplest integrable matrix model, the Ising matrix 
model, can be described by the Onsager algebra, which is a subalgebra of the cyclix algebra \cite{prl}.  As a 
result, all conserved quantities of the Ising model are elements of the cyclix algebra.  Likewise, we speculate 
that the Lie superalgebra describe the symmetry of supersymmetric matrix models, and may even help us determine the
integrability of them.

The support in part provided by the U.S. Department of Energy under grant DE-FG02-91ER40685 to us is acknowledged.

\end{document}